\documentclass{aa}

\begin{document}
	
\def\ref{\par\noindent\hangindent 20 pt}
\font\aut=cmbx10
\def\mincir{\ \raise -2.truept\hbox{\rlap{\hbox{$\sim$}}\raise5.truept  %MC
\hbox{$<$}\ }}                                                          %
\def\magcir{\ \raise -2.truept\hbox{\rlap{\hbox{$\sim$}}\raise5.truept  %
\hbox{$>$}\ }}                                                          %
%    Definition of references international style
% ----------------------------------------------------------------------------
%
\def\ea{{ et al. }}
\def\asec{$^{\prime\prime}$ \ }
\def\magsec{mag/$\Box^{\prime\prime}$ }	% mag/square_arcsec 
\def\cosmo{H$_0$ = 50~km~s$^{-1}$~Mpc$^{-1}$ and q$_0 = 0$} 
\thesaurus{03(11.01.2; 11.02.1; 11.09.2; 11.14.1; 11.16.1; 11.19.6)}

\input{psfig.tex}
\title { Optical imaging and spectroscopy of BL Lac objects. }

\author{Renato Falomo\inst{1} \and Marie-Helene Ulrich\inst{2} }
\institute{Osservatorio Astronomico di Padova, Vicolo dell'Osservatorio 5, 
35122 Padova, Italy; e-mail: falomo@pd.astro.it
\and ESO, Karl-Schwarzschild-Str. 2, D-85748 Garching bei Munchen, Germany;
    e-mail: mhulrich@eso.org}                                    
                                  
%\baselineskip=18pt

%VERSION 21 Feb 2000

\offprints{R. Falomo \\ 
{\it Based on observations collected at the European Southern Observatory,
La Silla, Chile.} }

\date{Received / Accepted}
\titlerunning{Imaging and spectroscopy BL Lacs}
\authorrunning{Falomo \& Ulrich}
\markboth{Falomo \& Ulrich: Imaging and spectroscopy of BL Lacs}{}
\maketitle

\begin{abstract}

We present optical images and spectroscopy for a dozen of BL Lac
objects. Most of these objects were not previously studied and we give
for the first time the properties of their host galaxies.  The
properties of the new host galaxies are generally consistent with
those derived in previous optical studies.  We found a case (1101-23)
where the external isophotes of the galaxy are clearly boxy.
 
In addition we gathered spectroscopy for several BL Lac objects with
unknown redshift and for companion galaxies.  This allowed us to
derive a tentative redshift for two new BL Lacs and to investigate the
environment around PKS 0829+04.  These data complement existing data
available in the literature on host galaxies of BL Lacs and their
(close) environments.


\end{abstract}

\section{Introduction}

In the past decade BL Lac objects have been actively investigated in
direct imaging and spectroscopy using ground based telescopes and
HST.The imaging effort has been directed towards detecting the host
galaxy, and when possible towards measuring its absolute luminosity
and colors and determining its morphological properties.  The aim of
the spectroscopy has been to measure the redshift of the host or to
measure the redshift of companions galaxies in order to assess a
possible group or cluster membership.

Apart from studies on individual objects a number of papers have
presented optical images for samples or lists of objects.

Twenty three objects have been imaged with the William Herschel
Telescope in the R filter and 14 are resolved (Abraham \ea1991).
However due to either unknown redshift or poorly detected nebulosity
only for 6 sources absolute quantities are derived. Some cases of disc
dominated host galaxies are proposed.

Sixteen objects in the southern sky have been studied by Falomo (1996)
using sub-arcsec images obtained at the ESO 3.5m New Technology
Telescope (NTT). Eleven sources were resolved and the hosts found to be
luminous ellipticals (M$_R \sim$ --23.5). For a number of objects close
companion galaxies are detected.  Due to their small projected
distance it is likely that they are associated with the BL Lac but
spectroscopy is needed to assess this point.

A larger sample but with poorer average resolution was investigated
using the 3.6m CFHT (Wurtz et al. 1996).  Fifty objects have been
observed and 36 well resolved. For another ten objects the host galaxy
has been only marginally detected. No difference of host properties is
found between objects discovered in radio surveys (i.e. 1Jy sample) and
those derived from X-ray surveys (i.e. EMSS). With very few exceptions
all the BL Lac objects investigated are classified as ellipticals
based on the surface brightness profiles.

More recently a study of the host galaxies in a large sample of X-ray
selected (high frequency peaked) BL Lacs have been presented by 
(Falomo \& Kotilainen 1999). They used high resolution images in the R
filter at the Nordic Optical Telescope (NOT) to image 52 targets from
EMSS and Einstein Slew samples.  All the 45 objects resolved are well
represented by elliptical models. On average the hosts are found 1
magnitude more luminous than M$^*$ (M$^*_R \sim$ -22.5; Mobasher et
al. 1993; assuming R-K = 2.7).  

In addition to ground based studies several 0.1 arcsec resolution
short exposure images have been obtained with WFPC2 camera on board of
HST during a snapshot survey (Scarpa \ea  2000; Urry \ea 2000). Objects
from various samples, and in the redshift interval 0.05 $< z <$ 1.3, 
 were observed and 69 out of 110 observed are resolved.
The highest redshift host galaxy detected is  $z=0.664$ for 
1823+568. For 80\% of the resolved host galaxies an elliptical 
model is clearly preferred over a disc galaxy.
The median absolute magnitude of these host  galaxies 
($M_R \sim -23.7$) is at least one magnitude brighter than $M^*$. 
The nuclei are always well centered over the body of the galaxy and 
have luminosity similar  to that of its host galaxy.
From the point of view of the optical morphology the hosts of BL Lacs 
appear indistinguishable from ``normal'' (non active) ellipticals. 

The main aim of all these observations outlined above was to detect
the host galaxies and to determine their structural and photometric
properties.  The knowledge of the kind of galaxies that host a BL Lac
phenomenon in the nucleus is of importance not only for 
understanding/studying the nuclear activity vs galaxy connection (see
e.g.  Lawrence 1999) but also as a probe to test unified models of
radio loud AGN. In particular if BL Lacs are FR I radio galaxies whose
jet is aligned along the line of sight (e.g. Urry \& Padovani 1995;
Ulrich 1989) their host galaxies should exhibit exactly the same
photometrical and morphological properties as the hosts of FR I. The
properties of the BL Lacs hosts can also be compared with those of
related beamed objects such as FSRQ and HPQ (see e.g. Kotilainen \ea 
1998a).

The aim of this work is to complement the existing data on BL Lac host
galaxies and close environment with new imaging and spectroscopy for a
dozen of (previously not well studied) objects.  A general discussion
and comparison of the properties of BL Lacs and radio galaxies will be
presented elsewhere.

In this paper we therefore present results from optical images of BL
Lac objects collected at the NTT with mostly sub-arcsec
resolution. Most of the objects presented here were not previously
investigated with adequate capabilities. These observations therefore
complement the existing data on BL Lac host galaxies.

We also present spectroscopic observations for some of the objects
performed with the aim of deriving the redshift of the host galaxies
and of some nearby companion galaxies.  When no spectroscopic redshift
is available we give an estimate of the photometric redshift derived
by assuming that the host has M$_R$ = --23.85 and R$_e$ = 9 kpc (the
typical median values found in previous studies of BL Lacs hosts;
e.g. Falomo \& Kotilainen 1999).

In Sect. 2 we describe the observations and data analysis. 
Section 3 reports the results obtained for each individual objects.
Section 4 gives a summary of the results and discussion.

\section{Observations and data analysis}

Optical observations were obtained using the 3.5m New Technology
Telescope (NTT) at the European Southern Observatory (ESO), operated
via remote control from the ESO headquarters in Garching (Germany).
We acquired images using the Superb Seeing Imager (SUSI; Melnick \ea
1992) which is installed at one of the Nasmyth foci of the NTT.
Configuration used was R-band filter and a CCD (TK 1024) with 24$\mu$m
pixel size corresponding to 0.13\asec on the sky.  Conditions were
photometric and seeing was ranging from 0.55 to 1.2 arcsec (FWHM), and
in most cases $<$ 1\asec. Observations of standard stars (Landolt
1992) were used to set the photometric zero point.

We obtained images centered on the BL Lac object with exposure times
ranging from 10 to 30 min (see Table 1). 
For many objects we also secured one short (2 minutes) 
 exposure in order to be sure to  get unsaturated images of the nucleus
of the targets and to enable us to use bright stars in the field to
study the PSF.

The images were processed in the standard way (bias subtracted,
trimmed, flat fielded, and cleaned of cosmic rays) using the Image
Reduction and Analysis Facility (IRAF) procedures.  A journal of the
observations is given in Table 1.

%Table 1 --------------------------------------------------
\begin{table*}
\begin{center}
%\caption{ Properties of the BL Lacs and journal of the observations.}
\begin{tabular}{lllllll}
\multicolumn{7}{c}{\bf Table 1 -- Journal of the observations.}\\
\hline\\
Name & z & Class$^{a)}$ & Date & T(int) &  Seeing & A$_R$ \\
     &   &       &      & (sec)  &  (arcsec) &          \\
\multicolumn{7}{c}{ SUSI Imaging R filter} \\
PKS 0138-097 &    0.733 &  LBL & 1996 Jan 19  & 600 &  1.2 & 0.14    \\
     0301-243 &     0.26 &  HBL & 1996 Jan 19  &1200 &  0.8 & 0.10   \\
     0338-21 &  (0.45) &  LBL & 1996 Jan 19  & 600 &  0.6 &  0.12  \\
 REX 0353-36 &  (0.40) &  HBL & 1996 Jan 19  & 600 &  0.6 &  0.04  \\
 PKS 0735+178 & $>$0.424 &  LBL & 1996 Jan 19  &1800 &  0.7 & 0.26   \\
 PKS 0736+017 &    0.191 & FSRQ & 1996 Jan 19  &1200 & 0.55 & 0.72   \\
 PKS 0818-128 &     ...  &  LBL & 1996 Jan 19  & 600 &  0.7 & 0.36    \\
   H 1101-23 &    0.186 &  HBL & 1996 Jan 19  &1200 &  0.9 & 0.24   \\
  MS 1312.1-422 &    0.108 &  HBL & 1996 Jan 19  & 600 &  1.4 & 0.36   \\
  MS 1332.6-293 &     0.25 &  HBL & 1996 Jan 19  & 600 &  1.5 & 0.22   \\
\hline\\
\multicolumn{7}{c}{ EMMI spectroscopy  } \\
	Name & z      & Slit/PA$^{b)}$  & Date       & T(int) &   & \\
\hline\\
     0301-243 & 0.26   & 2off/90  &1996 Jan 20 &3600    &   &  \\
 PKS 0548-322 & 0.068  & 2/90     &1996 Jan 20 &3600    &   &  \\
 PKS 0754+10 & 0.28   & 2off/0   &1996 Jan 20 &3600    &   &  \\
 PKS 0818-128 & ...    & 2/0      &1996 Jan 20 &1800    &   &  \\
 PKS 0829+046 & 0.18   & 2off/109 &1996 Jan 20 &3600    &   &  \\
%     1101-23 & 0.186  & 2/35     &1996 Jan 20 &1200   &   &  \\ 
\hline\\
\multicolumn{7}{l}{ $^{a)}$ LBL = Low frequency-peaked BL Lacs;  HBL =
High frequency-peaked BL Lacs;}\\
\multicolumn{7}{l}{ FSRQ = Flat Spectrum Radio Quasars}\\
\multicolumn{7}{l}{ $^{b)}$ Slit width in arcsec and position angle (degrees). 
off = offset from the nucleus by $\sim$ 1\asec }\\
\end{tabular}
\end{center}
\end{table*}
%\normalsize

Spectroscopy of the the objects and/or of galaxies in the field 
were obtained for some targets in order to
determine the redshift of BL Lacs and/or nearby companion galaxies.
For this purpose the ESO multi mode instrument EMMI (Melnick \ea 1992) was used
with red arm and grism elements. In general the slit has been
oriented in order to obtain in a single observation both the BL Lac
object and one or more galaxies around the source.

All the images have been analyzed following the methods and procedure
described in Falomo (1996). In particular surface photometry analysis
was performed down to the surface brightness magnitude $\mu_R \sim$ 26
mag./arcsec$^2$ in order to derive the properties of the host
galaxies.  A fit of the radial brightness profile was performed
assuming a simple two model components: a point source plus a
elliptical galaxy described by a de Vaucouleurs law
\begin{displaymath}
I(r)=I_0exp\{-7.67[(r/r_e)^{1/4}-1]\}\end{displaymath} 
where $I(r)$ is the surface brightness and $r_e$ the effective radius.\\
 Also disc galaxies models were attempted but in no cases
they gave a better fit than the elliptical model. This is  consistent with
what was found in previous studies on a larger number of 
sources ( Falomo \& Kotilainen 1999; Urry \ea 1999; Scarpa  \ea 2000).

To obtain absolute quantities we applied correction for Galactic
extinction and redshift (K-correction). The former was determined
using the Bell Lab Survey of neutral hydrogen N$_H$ converted to
E$_{B-V}$ (Stark et al. 1992; Shull \& Van Steenberg 1985), while the
latter was computed from the model of Coleman \ea  (1980) for
elliptical galaxies.  Throughout this paper, \cosmo ~are adopted.

\section{Results }

In Fig. 1 we report the observed radial brightness profile of the
objects together with the best fit with the two components (point
source plus elliptical galaxy) for the objects resolved. Parameters of
the fit and absolute quantities for host galaxies and the nuclei are
given in Table 2. In this Table columns 4--8 we give the results from
this paper.  The redshift in column 2 is drawn from literature except
that for 0301-24 and two cases where a photometric redshift 
(given in parenthesis) is derived
from the observed host properties. In the following discussion
absolute quantities are given including corrections for galactic
extinction and redshift (K-correction).  Optical spectra of the BL
Lacs or companion galaxies are reported in Fig. 2 together with the
main identifications of observed spectral features.

\begin{table*}
\begin{center}
\begin{tabular}{lllllllll}
\multicolumn{9}{c}{\bf Table 2 -- Properties of host galaxies and nuclei.}\\
\hline\\
	(1)   &   (2) &   (3)     &   (4)       &   (5)  &   (6)  &
(7) & (8) \\
     Name     &   z$^{a)}$   &     K$_R$ &  R(nucleus) & R(host) &  M$_R$(nuc)&
		M$_R$(host) & R$_e$(kpc) &  resolved$^{b)}$ \\
\hline \\
  PKS0138-097 &  0.733 &   1.46 & 17.64 &     * & -26.39 &    *  &  *&  N \\ 
      0301-24 &   0.26 &   0.28 & 15.96 & 17.46 & -25.36 & -24.14 & 23.1 &  Y \\ 
      0338-21 & (0.45)&      * & 16.44 & 18.86 &     *  &    *  &  *   & Y+ \\ 
   REX0353-36 & (0.40)&      * & 18.05 & 18.92 &     *  &    *  &  *  & Y+ \\ 
   PKS0735+17 & $>$0.424 &      * & 15.22 &   *   &    *   &    *   &  * &    N \\ 
      0736+01 &  0.191 &   0.20 & 16.34 & 17.08 & -24.87 & -24.33 & 12.0 &  Y \\ 
      0818-12 &      ? &      * & 16.17 &   *   &    *   &   *    &  *	 &  N \\ 
      1101-23 &  0.186 &   0.19 & 16.80 & 16.41 & -23.87 & -24.45 & 22.3 &  Y \\ 
    MS1312-42 &  0.108 &   0.10 & 18.64 & 16.25 & -20.89 & -23.38 &  5.3 & Y+ \\ 
    MS1332-29 &   0.25 &   0.27 & 19.44 & 20.36 & -21.92 & -21.27 &  4.2 & Y+ \\ 
\hline\\
\multicolumn{9}{l}{$^{a)}$ Photometric redshifts are enclosed within
parenthesis (see text)} \\
\multicolumn{9}{l}{$^{b)}$ N = not resolved; Y = resolved; Y+ =
resolved (first detection) } \\
\end{tabular}
\end{center}
\end{table*}

% Fig. 1 radial profiles
\begin{figure*}
\hfill
{\bf Fig.~1.}~~Continue.
\end{figure*}

\begin{figure*}
%\psfig{file=fig_1b.ps,width=18.3cm,height=16cm}
%\resizebox{18.3cm}{22cm}{\includegraphics {ntt96_fig1b.ps }}
\hfill
\caption{The observed radial luminosity profiles of each BL Lac object 
(filled squares), superimposed to the fitted model (solid line)
consisting of the PSF (short-dashed line), de Vaucouleurs bulge
(medium-dashed line). In the cases of unresolved sources only the
scaled PSF profile is shown.}
\label{Fig1_2}
\end{figure*}

%--- FIG 2 ---------------------------
% Figure 2  spectra of gals
\begin{figure*}
\hfill
\caption{Optical spectra of  BL Lac objects and companion galaxies. 
Fluxes are in relative units. }

\label{Fig1_1}
\end{figure*}

\subsection{Comments for individual objects}
\underline{ PKS 0138-097  } 

This object was observed under 1.2\asec  seeing and it looks
unresolved. Heidt \ea  1996 have presented deep sub-arcsec images of
this source that indicate the presence of close companion objects.
These could be responsible for the intervening absorption system at z
= 0.501 (Stickel \ea  1993) seen in the spectrum of the BL Lac
object.  Our image was taken under relatively poor seeing but
nevertheless some evidence of the southern feature at $\sim$1.5\asec 
from the center of the source is present in our image.
This object has also been  imaged by HST and found to be unresolved
(Scarpa \ea  2000) but the presence of a companion galaxy at 1.5\asec
South from the nucleus is clearly apparent. 

Recent spectroscopy (Stocke \& Rector 1997) detects for the 
first time the emission-line redshift of z=0.733 based upon weak Mg II
and [O II] emission features. At this relatively high redshift our
image result is consistent with this object being in a luminous (not
detected) host galaxy at z = 0.733.

\underline{ 0301-243  } 

We took a 20 minute image under good seeing (~0.8\asec~) of this BL
Lac object that clearly shows an extended nebulosity (ellipticity
$\varepsilon$ = 0.3; $\epsilon$ = 1- $b/a$ ) with a complex close
environment (see Fig. 3).

The immediate region around the object is rich with faint galaxies and
there is a marked enhancement of the galaxy density within
$\sim$60\asec from the BL Lac object.

The spectra of two galaxies (G1 and G2; see Fig. 3) at $\sim$ 6\asec
and 20\asec \ from 0301-243 indicate that they are at $z = 0.263$
suggesting a cluster of galaxies of Abell richness class 0 might be
associated with the BL Lac source at this redshift (Pesce \ea  1995).
The radial profile is adequately well represented by a point source
plus the elliptical model while the fit with an exponential disk is
not acceptable.  Fig. 3 (right panel) shows the field after
subtraction of the BL Lac model (nucleus plus host galaxy) revealing
the faint galaxy $\sim$ 3.5\asec \ South of the nucleus.  After
masking out the companion from the image we find that the surrounding
nebulosity is very well centered on the nucleus within an accuracy of
0.2\asec.

We took three optical spectra of the nebulosity with the slit off the
nucleus by 2\asec. They are still dominated by the signal from the
non-thermal source but all three show one weak emission line at
$\lambda$ = 6303 \AA \ (see Figure 2).  The most plausible
identification for this emission is [ O III] 5007 \AA \ that yields a
redshift of 0.26.  Fainter emissions like [ O III] 4959 \AA \ or H$_\beta$ 
could be present at this $z$ but not detectable in our spectrum because 
the features are lost in the noise.
Other possible identifications like MgII 2800 (at z = 2.25) 
are not acceptable because the host galaxy would be too luminous (M $<$ --30).
The redshift of 0.26 is very similar to the redshift of the
companion galaxies G1 and G2 (respectively of M$_R$ = -20.7 and M$_R$
= -22.3) and supports the idea that the host of the BL Lac is the
dominant member of a cluster of galaxies. We note that few other
examples have been reported in the literature of BL Lacs in clusters
whose membership has been proved spectroscopically.  
H0414+00 (Falomo \ea 1993a) 
is in a cluster of Abell class 0; PKS 0548-32 (Falomo
\ea  1995) is in a cluster of Abell class 1-2.

At this redshift (z = 0.26) the absolute magnitude of 
the host galaxy of 0301-24 is M$_R$ = -24.1

% Fig 0301-24 - contour
\begin{figure*}
\hfill
\caption{Contour plot of the central part of 
0301-24 before (left) and after (right) subtraction of a 2D model of the galaxy
and the  nucleus. After subtraction of the model a faint galaxy 3.5\asec South is visible together with some residual light from the bright nucleus. 
The spacing between isophotes is 0.5 magnitudes while
faintest shown level is $\mu_R$ = 25.5 mag arcsec$^{-2}$.
Galaxies G1 and G2 are at redshift z = 0.26.}
\end{figure*}

\underline{ 0338-21}

Our image obtained with 0.6\asec seeing shows the object to be
resolved.  This is the first detection of the nebulosity for the
source.  However its magnitude is not consistent with the redshift of
the object published twenty years ago ( z = 0.048 ; Wright \ea 1977).
The strongest absorption line identified in Wright \ea is indeed a
telluric band at 6870 \AA.  Subsequent spectroscopy of the source has
failed to confirm this redshift and a pure featureless optical
spectrum has been observed (Falomo \ea 1994). In fact at this redshift
of z = 0.048 the nebulosity would correspond to an unreasonably faint
and small host galaxy (M$_R \sim $ -18.5).  Assuming a typical host
galaxy (see Sect. 1) we can well fit the radial brightness profile
with a nucleus plus host galaxy obtaining a photometric redshift z
$\sim$ 0.45.

\underline{ REX 0353-36  } 

The source was identified as a BL Lac object in the REX survey of AGN
(Wolter \ea 1997). Its optical spectrum is featureless (Wolter \ea
1998). We obtained an image under very good seeing (0.6 arcsec) and are able
to detect the surrounding nebulosity and measure its luminosity and
R$_e$. This is the first detection of its host.  There is no
spectroscopic redshift but we can estimate a photometric redshift from
the image decomposition assuming the host galaxy has average
properties for BL Lacs hosts. The value of the photometric redshift so
obtained is z $\sim$ 0.4.

\underline{ PKS 0548-322}

This is a very well know BL Lac object at z = 0.068 (Fosbury \& Disney, 1976) 
with a very large host galaxy in a rich environment (Falomo et
al 1995). We took one relatively short exposure but good
signal-to-noise spectrum centered in the nucleus to search for
possible emission lines as have been reported in a number of nearby BL
Lacs (e.g. BL Lac itself, Vermeulen \ea 1995 ).The spectrum, shown
in Fig. 2, exhibits a substantial contribution from the stellar
population of the host galaxy. The MgI 5175 \AA \ and Na blend 5892 \AA \
are well detected with equivalent widths of 12 \AA \ and 6.5 \AA,
respectively.  We could not find any emission down to a limit of
equivalent with of 2 \AA. \ This limit corresponds to H$_\alpha$ line
luminosity L(H$_\alpha$) $\sim$ 5 $\times$ 10$^{40}$ erg s$^{-1}$
which is about a factor 10 lower than the line detected in BL Lac
(Vermeulen \ea 1995).

\underline{PKS 0735+178}

This BL Lac object is bright and strongly variable. It has been
extensively studied in the radio range and several moving components
have been detected in VLBI. The optical spectrum shows the absorption
line due to an intervening system at 3980 \AA, which if identified
with Mg II gives z $>$ 0.424 (Carswell \ea 1974) Our images were
obtained under seeing of 0.8\asec \ but the source remains
unresolved. Previous images were presented by Hutchings \ea (1988)
who also found this source unresolved.  There is no sign in our image
(see Fig. 4) that the galaxy 7\asec NW is distorted by interaction
with 0735+178 as suggested by previous lower resolution images
(Hutchings \ea 1988).  The object was also imaged by Stickel \ea
(1993) who are not able to detect the surrounding nebulosity. They
obtained a spectrum of the galaxy 7\asec NW and found z = 0.645.  
This BL Lac object is unresolved also in a short exposure
image obtained with HST (Scarpa \ea 2000).  In addition to the two
well resolved companion galaxies we detect a faint emission at $\sim$
3.5\asec East from the BL Lac (see Fig. 4).  Given its projected
distance from the BL Lac (25 kpc at z = 0.424) it could be related to
the intervening absorption at z = 0.424 but we cannot exclude that it
is just a faint background source.

From our image we can set a lower limit to the redshift (again
assuming the typical properties for the host) of z $>$ 0.5, 
consistent with the limit derived from intervening absorption. 

% Fig 3 :  0735+17 image
\begin{figure}
\hfill
\caption{The BL Lac object PKS 0735+17 (brightest object in the center) 
imaged by NTT+SUSI (R filter). The two bright objects at both sides of
0735+17 are stars. Field shown is 52 arcsec and North-East is at the top-left side.}
%\label{Fig0735}
\end{figure}

\underline{ PKS 0736+017}

The excellent (seeing 0.55 arcsec) image (see Fig. 1) shows the flat
spectrum radio quasar PKS 0736+01 (z = 0.191) as well as two close
resolved faint companions that are embedded in the nebulosity of the object.
The radial luminosity profile (see Fig. 1 ) is very well represented by
an elliptical galaxy with a bright point source in the nucleus. It is
found that the galaxy has M$_R$ = -24.3 and effective radius of $\sim$
12 kpc.

This host galaxy was previously detected in the optical with lower
resolution by Wright \ea (1998).  They derive M$_R$ = -22.0, which
is substantially fainter than our value. We note that this discrepancy
could be due to a problem in the Wright \ea image calibration as
their surface brightness goes unbelievably faint.  At 5\asec  from
the nucleus their surface brightness is about $\mu_R$ =28 while our
value at the same radius is $\mu_R$ = 24.

The object has been also resolved in the NIR by Taylor \ea (1996) 
who found M$_K$ = -26.3, and by Kotilainen \ea  (1998a) who found  M$_H$ =
-26.2.  The R-H color turns out to be $\sim$ 2.0, consistent
with the range of values reported by Kotilainen \ea 1998b for a
number of BL Lacs.

\underline{PKS  0754+10}

We took two spectra of this BL Lac object for which no firm value of
the redshift is available but whose host galaxy had already been
detected (Abraham \ea 1991; Falomo 1996).  The tentative redshift
($z=0.66$) proposed by Persic \& Salucci (1986) based on inspection of
the photographic spectrum reported by Wilkes \ea (1983) is unlikely
as the host galaxy would be extremely luminous (($M_R \sim -26$~mag).
Our spectra were obtained positioning the slit 2\asec from the nucleus
in order to reduce the contamination of light from the bright nucleus.
Therefore the spectrum (see Fig. 2) is noisy and it is still
dominated by the nuclear non thermal emission. We are not able to
unambiguously identify spectral lines but some hint of the CaII break
signature from the host galaxy is possibly apparent at $\lambda$ =
5045 \AA \ which corresponds to z = 0.28.  At this redshift the
detected surrounding nebulosity would be M$_R$ $\sim$ -23.  We note
that this is consistent with the value of the redshift of the
companion galaxy (see Fig. 5) 13.6\asec north-east of the BL Lac
object ($z$=0.27; Pesce \ea 1995) and could be another case of a
companion galaxy physically associated with a BL Lac object. A
definitive redshift determination is however still needed for this BL
Lac object.

% Fig 0754+70 image
\begin{figure}
\hfill
\caption{Image of the BL Lac object (brightest object in the center) 
PKS 0754+10 obtained with NTT+SUSI
(R filter). The galaxy G1 at 13\asec NE is at z = 0.27. Field shown is 52 arcsec and North-East is at the top-left corner.}
%\label{Fig0754}
\end{figure}

\underline{PKS 0818-128}

There is no redshift for this object and its optical spectrum is
featureless (Falomo \ea 1994).  Our optical images, obtained with
seeing of 0.7\asec, are not able to detect the host galaxy.  The
radial brightness profile is well matched by that of a scaled PSF (see
Fig. 1).  We can set a lower limit to the redshift assuming its
nucleus is hosted by a standard luminous ( M$_R$ $\sim$ -23.8)
elliptical.  The limit of redshift we found for such a galaxy to be
undetected in our image is z $>$ 0.5.

In order to search for emission or intervening absorption line we
gathered spectra in a wide wavelength range.  Our spectrum ( see
Fig. 2) is still dominated by the non-thermal featureless
emission. The only feature (in addition to telluric bands) we can
detect is an absorption at 6284 \AA \ (e.w. 0.7 \AA).  The most likely
identification of this feature is with an interstellar diffuse
absorption band at the same wavelength.  This is consistent with the
low galactic latitude (b$_{II} \sim$ 13$^{o}$) of the source.
Alternatively the absorption line could be identified with MgII 2800
\AA\ and this  would yield approximately z $>$ 1.2 and, 
consequentially,  the object would be extremely luminous (M$_R < --29$).  

\underline{PKS 0829+046}

Previous images obtained at sub-arcsec resolution showed that the host
galaxy (z = 0.18) has M$_R$ $\sim$ -23 (Falomo 1996).  There is also
an excess density of galaxies around this object (Pesce \ea
1994). But our spectroscopy shows that only some of them may be
physically associated with the BL Lac object.  Pesce \ea 1994
obtained the redshifts of galaxies G1 and G2 (see Fig. 6) at
respectively z =0.24 and z = 0.204.  We took additional spectra of
two other galaxies (G3 and G4, see Fig.s 2 and 6). We found that G4
is at significantly higher redshift (z = 0.29) while G3 is at z =
0.175, consistent with being associated with PKS 0829+04 at projected
distance of $\sim$ 120 kpc. In fact G3 is the only galaxy which is at
the same redshift as the BL Lac.  On one hand this is another case of
similar redshift of a companion galaxy and its BL Lac. On the other
hand the environment of 0829+04 must be less rich than what can be
estimated from galaxy counts.

% Fig 0829+04 image
\begin{figure}
\hfill
\caption{The field around the BL Lac object PKS 0829+04 obtained with NTT+SUSI
(R filter). Field shown is 2.2 arcmin and North-East is at the top-left corner.}
\end{figure}

\underline{H 1101-23}

This is a BL Lac discovered from X-ray survey and is surrounded by a
conspicuous rather elongated nebulosity (see Fig. 7 ) at the
proposed $z = 0.186$ (Remillard \ea 1989) confirmed by Falomo \ea
(1994).  The radial brightness profile extends to 15 arcsec along the
major axis.  We found that the luminosity profile is well fitted by an
elliptical galaxy model plus a point source.  The luminosity of the
host galaxy is very high. The absolute magnitude, M$_R$ = --24.45,
sets this galaxy among the brightest hosts of BL Lac objects (Falomo
\& Kotilainen 1999).

For this object (see Fig. 8) we performed detailed surface photometry
analysis using the AIAP package (Fasano 1990) in order to study the
structural properties of the galaxy.  From this analysis we derived
photometric and structural parameters (surface brightness,
ellipticity, position angle and Fourier coefficient C$_4$ describing
the deviation of isophotes from the ellipse) as a function of the
equivalent radius $r = a\times(1-\epsilon)^{1/2}$ where $a$ is the
semi-major axis and $\epsilon$ is the ellipticity of the ellipse
fitting a given isophote.  We found the ellipticity profile is
increasing from the center outwards up to $\epsilon$ = 0.45. The
profile of the C$_4$ (see Fig. 9) shows {\it disky} (positive C$_4$)
trend in the inner region while the external isophotes are
substantially {\it boxy} (negative C$_4$), possibly due to merging
processes ( e.g. Bender \ea 1988).  This is the only clear
evidence of significantly {\it boxy} isophotes ever found in a BL Lac
host.

Another example of a very luminous host galaxy (M$_{\rm R}$= --24.8, or
-24.45 if de Vaucouleurs law is fitted) was reported by Heidt \ea
(1999) for 1ES 1741+196. Also in this case the host galaxy isophotes
have high ellipticity ( $\sim$ 0.35). There is no information,
however, about the detailed shape of the isophotes and the amount of
possible boxiness.

\begin{figure}
\hfill
\caption{ Central portion (52\asec $\times$ 52\asec ) of the NTT + SUSI
image of the BL Lac object H 1101-23 (brightest object in the
center). A {\it squared} shape of isophotes are well apparent as
signature of boxiness. North is up and East to the left}
\end{figure}

% Fig. 3 %  1101-23 contour 
\begin{figure}
\hfill
\caption{Contour plot of 1101--23. 
Faintest isophote is $\mu_{\rm R}$ = 25.2
and spacing between isophotes is 0.5 mag. External isophotes 
are significantly boxy. }
\end{figure}

% fig 10 -- C4
\begin{figure}
\hfill
\caption{ The amplitude of the Fourier coefficient C$_4$/100 showing the
isophotes of the host  galaxy of H 1101-23 
being {\it disky} around 3-4 arcsec and 
substantially $boxy$ at larger radii. }

\end{figure}

\underline{MS 1312.1-422}

This source, drawn from the EMSS of BL Lacs (Maccacaro \ea. 1994) was
observed during bad seeing conditions (seeing of 1.4\asec) but since
it is at relatively low redshift ( z = 0.108; Morris \ea 1991) it is
rather well resolved. The host galaxy is indeed dominant with respect
to the nuclear source ( ratio nucleus/host = 0.1). Our fit of the
brightness profile yields M$_R$ = -23.4.  No other detection of this
host galaxy can be found in the literature.

Note that in the calculation of $\alpha_{OX}$ it is usually the
luminosity of the whole object (nucleus + host) which is used in the
calculation. Such a procedure if applied to 1312-42 would overestimate
the optical flux by a factor $\sim$ 10.

\underline{MS 1332.6-293}

The target belongs to the EMSS sample of BL Lacs although its
classification is uncertain. Optical spectra showed either emission
lines at z=0.256 or strong CaII break (Stocke \ea 1991 ) due to a
substantial contribution from stellar emission.

Our image shows this object is only marginally resolved. This is in
part due to the bad seeing ($\sim$ 1.5\asec) and also because the host
galaxy is substantially under-luminous (M$_R$ = -21.3) with respect to
the average of the host galaxies of BL Lacs (M$_R$ = -23.8 Falomo \&
Kotilainen 1999).

We note that the same object (1ES1322-297) is listed in the Einstein
Slew sample of BL Lacs (Perlman \ea 1996) and has a redshift z = 0.512
quite different from the previous finding.  Since in neither cases
there are spectra published we are not able to make our own judgment
of the validity of the redshift values.  However, the latter value
seems confirmed by another optical spectrum (albeit noisy) reported by
Rector (1998).  At z = 0.512 the host galaxy and point source would be
much more luminous (M$_{nuc}$ = -23.5 and M$_{host}$ = --23.3) and
well within the averages of these types of objects.

\section{Summary and conclusions}

We have presented optical images of a number of BL Lacs that were not
previously well studied. For several of these objects the first
detection of the host galaxy is presented here. The properties of the
hosts are consistent with them being luminous ellipticals as found in
previous similar studies. For two of the resolved objects that have
not a spectroscopic redshift we derive a photometric redshift based on
the observed properties of the surrounding nebulosity.
\bigskip

\noindent
{\it A bright and boxy elliptical}
\bigskip

We find that the external isophotes of the luminous host galaxy of
1101-23 are significantly {\it boxy} while the inner {\it disky}
region suggests the presence of a small disc component.  This is the
first clear example of a {\it boxy} galaxy hosting a BL Lac
object. Boxy isophotes are observed in a fraction of luminous
ellipticals (Bender 1988) and could be ascribed to merging events from
equal mass galaxies (e.g. Naab \ea 1999).  It would be interesting
to know what fraction of hosts of BL Lacs exhibit boxy isophotes as
compared with non active ellipticals.  Very little data are, however,
available on isophote shapes of BL Lacs host galaxies because the
presence of the bright nucleus and the quality of data often hinder a
reliable estimation of this parameter especially at high redshift.
For relatively low redshift objects with high resolution images it
should be possible to investigate the isophote shape in a systematic
way.
\bigskip

\noindent
{\it The immediate environment of BL Lacs}
\bigskip

Our spectroscopy has allowed us to derive a redshift for 0301-24 (z =
0.26) and possibly for 0754+10 (z=0.28). Both objects have companion
galaxies at redshifts very similar to that of the BL Lacs.  The
companions and the BL Lacs are thus very likely to be gravitationally
bound.  A third case is PKS 0829+04 for which we took the spectra of
two galaxies in the immediate environment and found that one is at the
same redshift as the BL Lac object. These spectroscopic results
improve the scanty data on redshifts of companion galaxies of BL
Lacs. Together with previous findings ( Falomo \ea 1993a,b; Pesce et
al 1994,1995; Heidt \ea 1999) our new results yield convincing
evidence that galaxies around BL Lacs are (often) gravitationally
bound with the BL Lacs. On the other hand only in very few cases do
these interactions lead to significantly (observable) disturbed
morphology (see e.g. Falomo \ea 1995, Heidt \ea 1999).

\begin{acknowledgements}

This work was partly supported by the Italian Ministry for University
and Research (MURST) under grant Cofin98-02-32.  This research has
made use of the NASA/IPAC Extragalactic Database (NED) which is
operated by the Jet Propulsion Laboratory, California Institute of
Technology, under contract with the National Aeronautics and Space
Administration.  We thank A. Wolter for providing coordinates and
finding charts of REX 0353-36 before publication.  RF thanks the ESO
visitor program for hospitality during several visits at the ESO
headquarter.

\end{acknowledgements}

\section{References}

\ref{} Abraham R.G., McHardy I.M., Crawford C.S. 1991, MNRAS 252, 482
\ref{} Bender R. 1988, A\&A 193, L7
\ref{} Bender R., Dobereiner S., Mollenhoff C. 1988, A\&AS 74,385.
\ref{} Carswell R.F., Strittmatter P.A., Williams R.E., Kinman T.D., 
         Serkowski K. 1974, ApJ 190, L101
\ref{} Coleman G.D., Wu C.C., Weedman D.W. 1980, ApJS 43, 393
\ref{} Falomo R. 1996, MNRAS 283, 241
\ref{} Falomo R.,  Kotilainen J. 1999 A\&A 321, 374
\ref{} Falomo  R., Pesce  J. E.,  Treves  A.  1993a,  AJ  105, 2031
\ref{} Falomo R., Pesce J.E.,  Treves A. 1993b ApJ 411, L63
\ref{} Falomo R., Scarpa R., Bersanelli, M. 1994 ApJS 93, 125
\ref{} Falomo R., Pesce J.E.,  Treves A. 1995 ApJ 438, L9
\ref{} Fasano G., 1990, Internal report of Astr. Obs. of Padova
\ref{} Fosbury R.A.E., Disney M.J. 1976, ApJ 207, L75
\ref{} Heidt J., Nilsson K., Pursimo T., Takalo,L.O., Sillanp\"a\"a A. 1996, A\&A 312, L13
\ref{} Heidt  J., Nilsson  K., Fried J. W., Takalo L. O., Sillanp\"a\"a A. 1999, A\&A 348, 113
\ref{} Hutchings J.B., Johnson  I., Pyke  R. 1988, ApJSS  66, 361
\ref{} Kotilainen J.K., Falomo R., Scarpa R. 1998a, A\&A 332, 503
\ref{} Kotilainen J.K., Falomo R., Scarpa R. 1998b, A\&A 336, 479
\ref{} Landolt A.U. 1992, AJ 104, 340
\ref{} Lawrence A. 1999, Adv. Space.Res  23, 1167
\ref{} Melnick J., Dekker H., D'Odorico S. 1992, The EMMI and SUSI ESO Operating Manual
\ref{} Maccacaro T., Wolter A., McLean B., et al. 1994,
		Astroph. Lett. Comm. 29, 267
\ref{} Mobasher B., Sharples  R.M., Ellis  R.S. 1993, MNRAS 263, 560
\ref{} Morris S.L., Stocke J.T., Gioia I.M., \ea 1991, ApJ 380, 49
\ref{} Naab T., Burkert A., Hernquist L., 1999 ApJ 523, L133
\ref{} Perlman E.S., Stocke J.T., Schachter J.F., et al. 1996, ApJS 104, 251 (P96)
\ref{} Persic M.  Salucci P. 1986, in {\it Structure and Evolution 
	of Active Galactic Nuclei}, ed. G. Giuricin, F. Mardirossian, 
	M. Mezzetti M. Ramella, p. 657
\ref{} Pesce J. E., Falomo R., Treves A. 1994 AJ, 107, 494
\ref{} Pesce J. E., Falomo R., Treves A. 1995 AJ, 110, 1554
\ref{} Rector T., 1998, PhD. Thesis
\ref{} Remillard R. A., et al 1989, ApJ 345, 140 
\ref{} Scarpa R.,Urry C.M., Falomo R., Pesce J.E., Treves A. 2000, ApJS in press
\ref{} Shull J.M., Van Steenberg M.E. 1985, ApJ 294, 599
\ref{} Stark A.A., Gammie C.F., Wilson R.W., et al. 1992, ApJS 79, 77
\ref{} Stickel M., Fried J.W., Kuhr H. 1993, A\&AS 98,393
\ref{} Stocke  J.T.,  Rector  T.A. 1997, ApJ 489, L17
\ref{} Stocke  J.T., Morris S.L., Gioia I.M., et al 
%Maccacaro T., 	Schild R., Wolter A., Fleming T. A., H. J. Patrick 
1991 ApJS, 76 813
\ref{} Taylor G. L., Dunlop J. S., Hughes D. H.,  Robson E. I.
        1996, MNRAS 283, 930
\ref{} Ulrich M.H. 1989, in BL Lac Objects, ed. L. Maraschi,
	T. Maccacaro,  M. H. Ulrich, p 45
\ref{} Urry C.M.,  Padovani P. 1995, PASP 107, 803
\ref{} Urry C.M., Falomo R.,  Pesce J., Scarpa R., Treves A., Giavalisco M. 1999 Ap. J, 512, 88.
\ref{} Urry C.M., Scarpa R., O'Dowd M., \ea 2000  ApJ in press
\ref{} Vermeulen R.C. et al 1995 Ap, 452, L5
\ref{} Wilkes B.J., Wright A.E.,  Jauncey D.L.  Peterson B.A. 1983, 
		Proc. Astron. Soc. Aust. 5, 2
\ref{} Wright A.E., Jauncey, D.L. Peterson, B. A. Condon, J.J 1977 Ap.J 211, L115
\ref{} Wright S.C., McHardy I.M., Abraham R.G., 1998 MNRAS 295, 799
\ref{} Wurtz R., Stoke J.T., Yee H.K.C. 1996, ApJS 103, 109
\ref{} Wolter A., Ciliegi P., della Ceca R., et al 
% Gioia I.M., Giommi P., Henry J.P., Maccacaro T., Padovani P.,
% Ruscica C. 
1997 MNRAS 284, 225
\ref{} Wolter A.,  Ruscica C., Caccianiga A. 1998 MNRAS 299, 1047

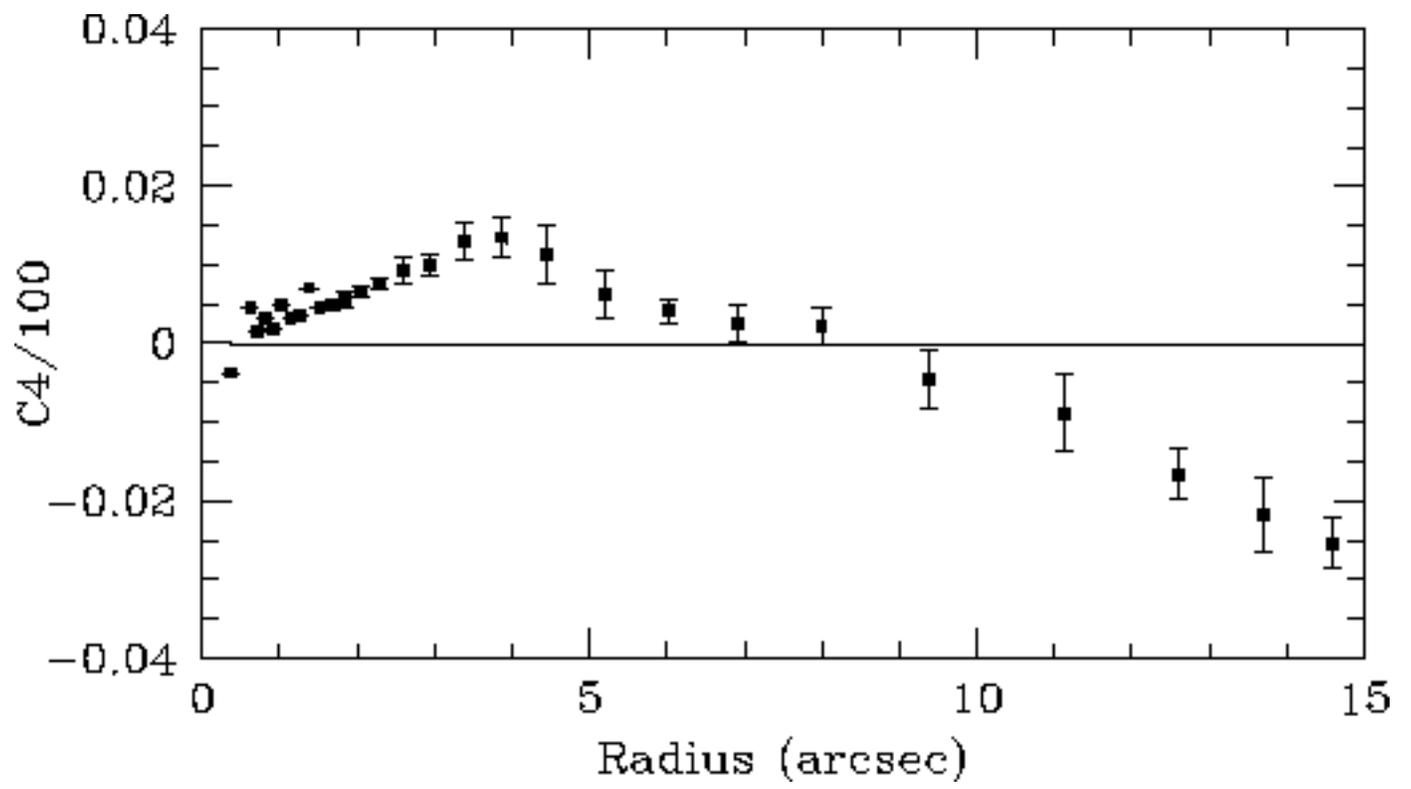
\end{document}